\documentclass[a4paper]{jpconf}
\usepackage{graphicx,pstricks,epsfig}
\begin{document}
\title{BCS-BEC crossover in spatially modulated fermionic condensates}

\author{Armen Sedrakian}

\address{Institute for Theoretical Physics, 
  J. W. Goethe-University, \\
Max-von-Laue-Str. 1, 60438 Frankfurt am Main, Germany
}

\ead{sedrakian@th.physik.uni-frankfurt.de}

\begin{abstract}
  Several novel multi-component fermionic condensates show universal
  behavior under imbalance in the number of fermionic species. Here I
  discuss their phase structure, thermodynamics, and the transition
  from the weak (BCS) to strong (BEC) coupling regime. The
  inhomogeneous superconducting phases are illustrated on the example
  of the Fulde-Ferrell phase which appears in the weak coupling regime,
  at low temperatures and large asymmetries. The inhomogeneous
  phases persist through the crossover up to (and possibly beyond) the
  transition to the strong coupling regime.
\end{abstract}

\section{Introduction}
The last decade has seen  impressive advances in the research, both
theoretical and experimental, on pairing in novel fermionic systems
which share a number of common features. One example is
encountered in experiments with ultracold atomic vapors in magnetic
traps. The atomic gases allow for a remarkable control over the parameter
space, including variations in the strength of the pairing force via
Feshbach resonance. The second example of interest is the nuclear/quark
matter which can be either spin polarized (by strong magnetic fields)
or isospin polarized, as required by the beta equilibrium in the
interiors of compact stars.

The possibility of transition from Bardeen-Cooper-Schrieffer (BCS)
pairing to Bose-Einstein condensation (BEC) in ensembles of attractive
degenerate fermions was conjectured long ago~\cite{NSR}. This
transition takes place when the dimensionless coupling $\lambda$,
which is a product of the density of states at the Fermi surface and
the dimensional coupling of the theory increases from weak coupling
($\lambda \ll 1$) to strong coupling ($\lambda \gg 1$).
Experimentally, it is achieved by variations of the coupling constant
via the mechanism of Feshbach resonance in atomic gases. For particles
interacting via the strong force the interactions cannot be varied at
will, but the range over which they are effective and the density of
states at the Fermi surface will change with density.  Therefore,
the combined effect of density variation on the density of states and
effective range of the coupling can lead to a crossover in strongly
interacting systems with density gradients.

The generalization of the Nozi\`eres-Schmitt-Rink (NSR) theory to
ensembles with population imbalance was carried out in the context of
the strongly interacting matter in ref.~\cite{LNSSS}. Experimental
realizations in the atomic vapors were achieved somewhat
later~\cite{Jin}.  The novel feature of the Hamiltonian describing
these imbalanced gases or (super)fluids is the non-invariance under
exchange of paired particles, i.e., there is a new scale in the
problem $\delta \mu = (\mu_{\alpha}-\mu_{\beta})/2$, where
$\mu_{\alpha}$ and $\mu_{\beta}$ are the chemical potentials of the
species. Furthermore, for certain values of asymmetry $\delta\mu$ the
system may support a current carrying state, where the
counter-propagating super-current and normal current cancel each
other. This phase is known as the Larkin-Ovchinnikov-Fulde-Ferrell
(LOFF) phase~\cite{LO,FF}. The LOFF condensate is spatially modulated,
therefore, minimally, it breaks the $O(3)$ rotational symmetry of the
system down to $O(2)$. However, in general, the condensate may assume
a complicated structure which breaks also translational symmetries.
The emergence of the LOFF phase becomes energetically favorable for
the following reason: setting up the condensate and the normal
excitations in motion costs kinetic energy, which is always positive
and thus unfavorable; however, the current changes the way the
phase-space is sampled in the vicinity of the Fermi surface, which in
turn enhances the condensation energy. The interplay of these two
effects tells us whether the LOFF phase is favored or not. Indeed, in
the low-temperature and large asymmetry sector of the phase diagram
the LOFF phase is favored over the BCS state. The way the LOFF phase
evolves into the BEC regime is not fully understood. I shall provide
some answers to this problem below. An alternative to the LOFF phase,
is the phase which allows deformations of the Fermi surfaces of the
species~\cite{DFS1}. The emergence of deformed Fermi surface pairing
(DFSP) can be explained in full analogy to the LOFF phase and I will
not repeat the arguments above.
\begin{figure}
\begin{center}
{\includegraphics[width=10cm,height=7cm]{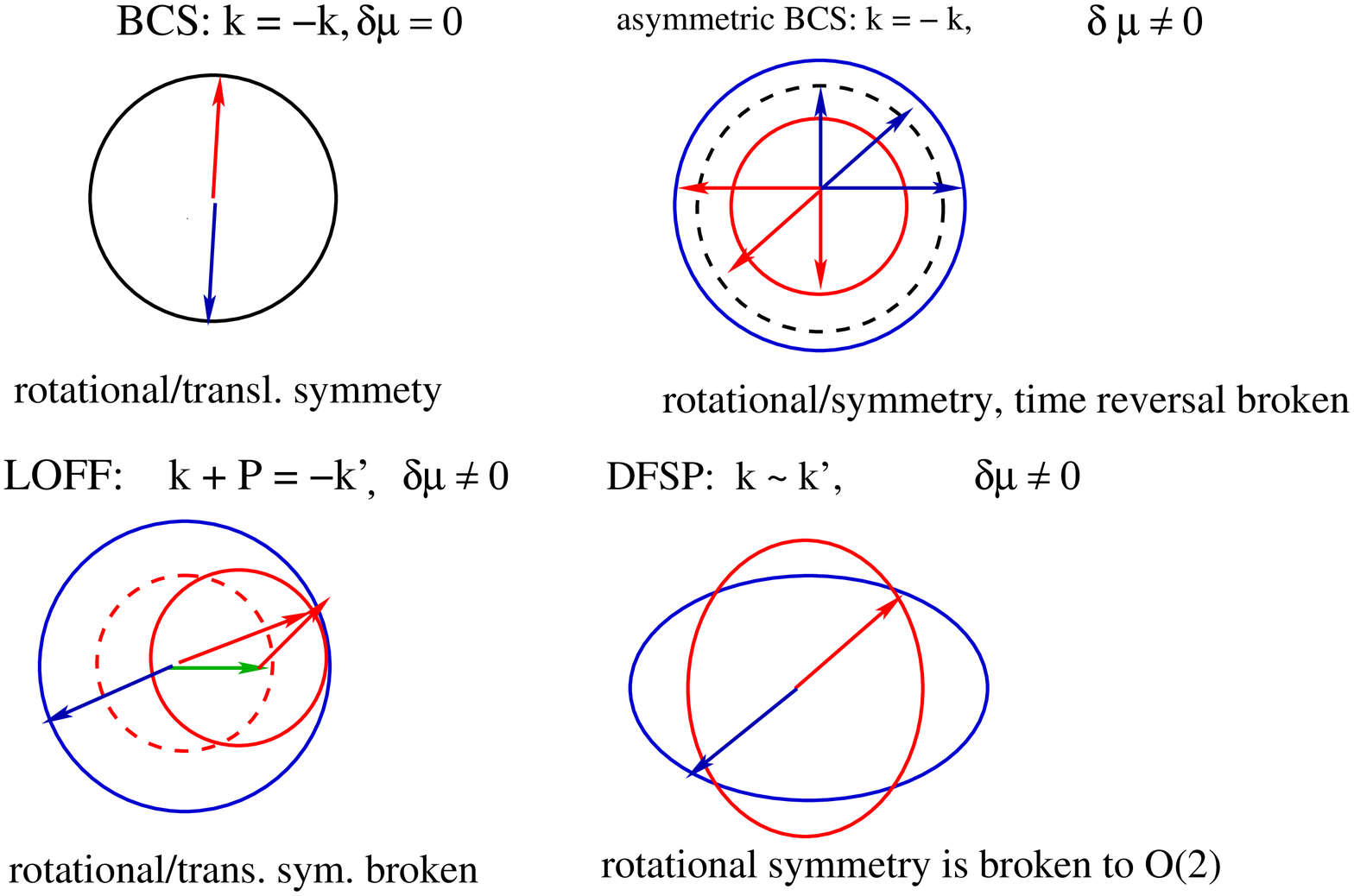}}
\caption{\label{fig:1} Schematic illustration of the Fermi spheres of
  two components (solid - majority component, dashed - minority
  component) and the momenta of paired fermions for four phases: BCS,
  asymmetric BCS, LOFF and DFSP.}
\end{center}
\end{figure}

\section{Homogeneous phases}

Let us start our discussion with the homogeneous phases, i.e., the
ordinary BCS phase and the asymmetrical BCS phase.  Once the pairing
interaction is specified, the mean-field solutions are obtained by
solving the gap equation together with the equations for the densities
for each fermionic component (see, e.~g., refs.~\cite{DFS1,DFS2}). The
solutions of the coupled integral equations then provide the value of
the gap parameter for fixed net density and asymmetry in the
populations of fermions measured either in terms of the density
asymmetry (hereafter $\alpha$) or the shift in the chemical potentials
$\delta\mu$.
\begin{figure}[h]
\begin{center}
\begin{minipage}{14pc}
\includegraphics[width=14pc]{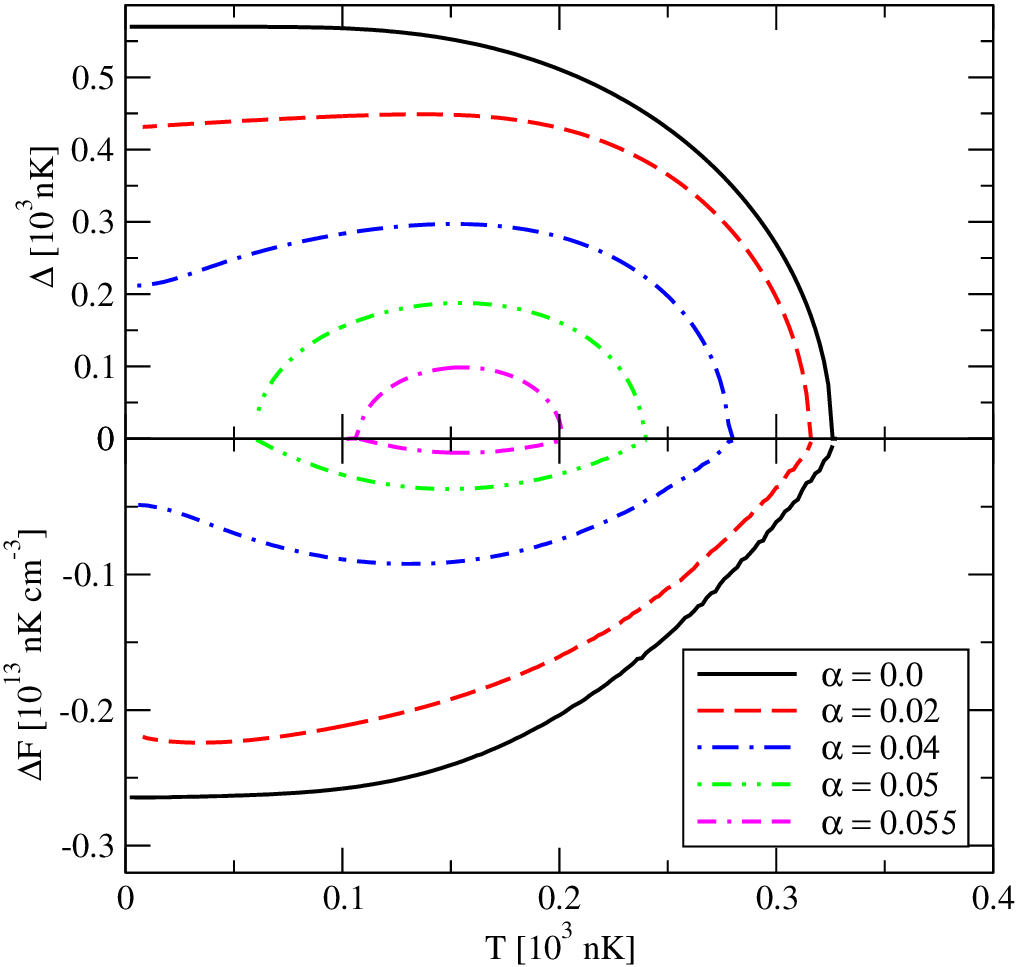}
\caption{\label{fig2} Dependence of the gap (upper panel) and the free energy
  (lower panel) on the temperature for an ultracold fermionic gas
  (Li$^6$) for several asymmetries.}
\end{minipage}\hspace{2pc}%
\begin{minipage}{14pc}
\includegraphics[width=14pc]{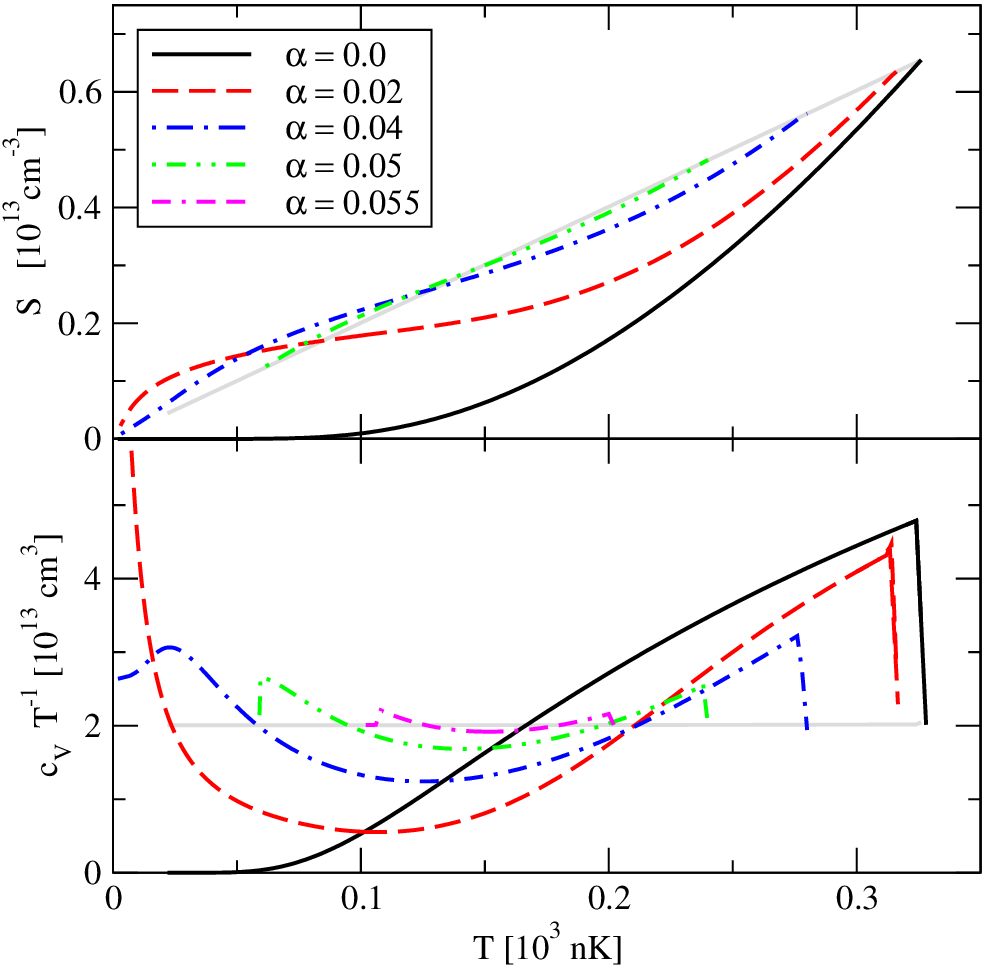}
\caption{\label{fig3} Same as in Fig. \ref{fig2}, but for the entropy
  (upper panel) and the specific heat (lower panel) for several asymmetries. }
\end{minipage} 
\end{center}
\end{figure}
\begin{figure}[h]                          
\vskip 0.3 cm
\begin{center}                                     
\includegraphics[width=14pc,height=12pc]{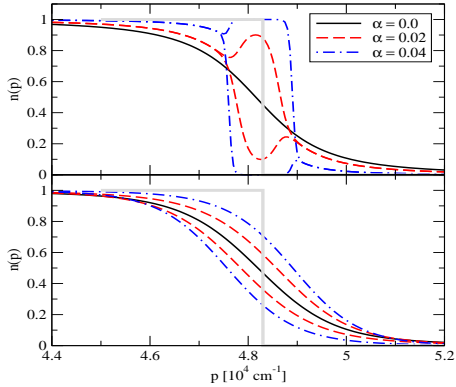}\hspace{2pc}%
\begin{minipage}[b]{14pc}
  \caption{\label{fig4} The occupation numbers of two components for
    various asymmetries.  Upper panel corresponds to the low
    temperature limit, the lower panel - to the high temperature
    limit. The minority component is extinct close to its Fermi
    surface for large asymmetries and  low temperatures. This effect
    is ``smeared out'' at higher temperatures (lower panel). }
\end{minipage}
\end{center}                                                                  
\end{figure}   
Figures \ref{fig2} and \ref{fig3} show the temperature dependence of
various quantities for fixed density
asymmetries~\cite{Sedrakian:2006mt}. The pairing correlations are
clearly suppressed for large asymmetries; in the high-temperature
domain $T\to T_c$ the temperature dependence of the functions is
analogous to the BCS theory for balanced systems. The low temperature
domain $T\to 0$, on the contrary, is anomalous, i.e., all the quantities
show temperature dependence that is different from the BCS theory. The
reduction of the gap, for example, is the result of the loss of
coherence between the paired fermions (or equivalently, overlap
between the thermal bands around the Fermi surfaces). For large
asymmetries this leads to a complete disappearance of the gap in the
limit of $T\to 0$, which manifests itself in the anomalous jump of the
specific heat (Fig.~\ref{fig3}).  While such a jump could be a
signature of the asymmetrical BCS state, we will see below that the
temperature anomalies are removed if one allows for the LOFF (or DFSP)
phase. This, of course, does not exclude the possibility of observing
the second jump in the specific heat if the gas is prepared at
high temperatures, where the BCS phase is stable, and is cooled down
into the low temperature unstable phase~\cite{Sedrakian:2006mt}.

Another interesting observation is the emergence of the empty shell
around the Fermi surface of the minority component for large
asymmetries seen in Fig.~\ref{fig4}. This leads to the notion of
``breached pairing'' superconductivity~\cite{Liu,Forbes}. Clearly the topology of the 
minority Fermi surface changes as the asymmetry is increased. However, the
topological phase transition disappears in the high temperature regime
(Fig.~\ref{fig4} lower panel).

Next we turn to the question on how the Nozi\`eres--Schmitt-Rink 
theory is modified in asymmetrical systems. This problem was first
studied in the context of pairing in low-density nuclear matter in the
attractive $^3S_1$-$^3D_1$ coupled channel, where neutron-proton Cooper pairs
in the BCS limit transform to a BEC of deuterons~\cite{LNSSS}. Following the NSR
conjecture one can directly extrapolate the BCS equations in the
strong coupling regime to obtain the properties of the deuteron
condensate. The pair wave-function, which is defined as the correlation
function of two fermionic creation operators $\psi(k) = \langle
a_{\vec k}^{\dagger} b_{-\vec k}^{\dagger} \rangle$, where
$a^{\dagger}$ and $b^{\dagger}$ are the creation operators of neutrons and
protons, obeys a Schr\"odinger type equation with an eigenvalue
equal $2\mu$, with $\mu$ being the chemical potential in the
symmetrical case.
\begin{figure}
\begin{center}
{\includegraphics[width=8cm,height=12cm,angle=-90]{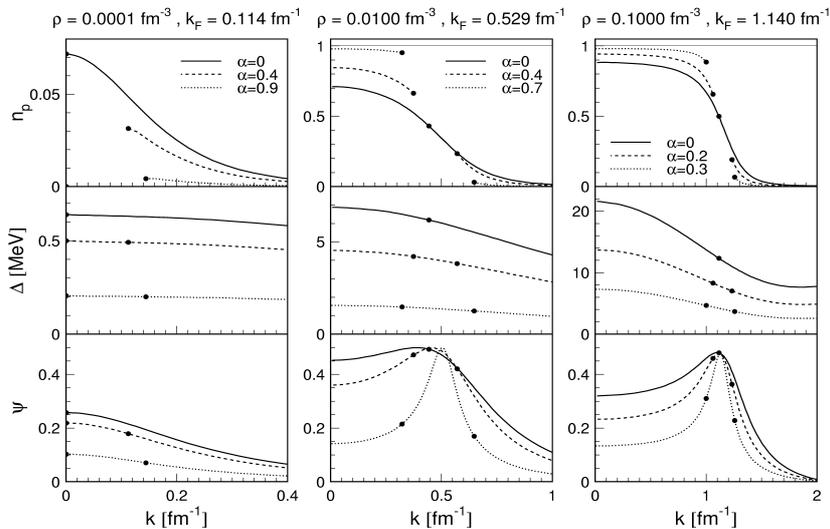}}
\caption{\label{fig5} BCS (right column) to BEC (left column)
  crossover in an imbalanced system. Upper
  panel: the occupation number of minority component; middle panel: the pairing
  field; lower panel: the condensate wave-function. Figure taken from ref.~\cite{LNSSS}.}
\end{center}
\end{figure}
The state that emerges in the strong coupling limit corresponds to a
mixture of deuterons (strongly bound two-body states) plus a fluid of
unbound neutrons at arbitrary large asymmetries.  At low densities the
Pauli-blocking plays a minor role and its modifications to the
deuteron wave-function are marginal. As a consequence large
asymmetries are compatible with the existence of a deuteron
condensate. Figure~\ref{fig5} shows the evolution of occupation
numbers of the minority component (upper panel), pairing gap
(intermediate panel) and condensate wave-function (lower panel) across
the BCS-BEC transition (from left to right). In this particular case,
the transition is enforced via the reduction of the net density.  A
remarkable feature of the transition is the change in the occupation
numbers of the minority component: the gap (breach) is widened across
the transition and in the BEC regime an empty (unoccupied) sphere
emerges inside the Fermi sphere. Clearly, the topology of the
Fermi sphere changes from having a strip of empty states to a empty
interior region; thus, the BCS-BEC transition is associated with a
topological phase transition in the shape of the minority Fermi sphere
at non-zero asymmetries. The BCS vs BEC nature of the condensate is
most clearly seen in the form of the condensate wave-function. In the
momentum space it is strongly peaked in the BCS limit and is flat in
the BEC limit (Fig.~\ref{fig5} lower panel). In the real space this
picture translates into loosely bound state whose wave-function
oscillates over many periods in the BCS limit. In the opposite BEC
limit one finds tightly bound state with a strongly localized 
wave-function.

\section{Inhomogeneous phases}
In this section I will explore the modifications that arise when
inhomogeneous phases are allowed to exist along with the symmetrical
and asymmetrical BCS phases.  I will illustrate the main points on the
example of color superconductivity in cold quark matter with pairing
among the lightest quarks of different flavors ($u$ and $d$) in the
color anti-triplet state (see
e.g. refs. \cite{BR02,Casalbuoni:2003jn}). Our Ansatz for the
inhomogeneous condensate corresponds to the Fulde-Ferrell (FF) phase
which postulates a single plane-wave dependence of the gap on the
total momentum $\Delta (Q) \propto \Delta_0
\exp(iQr)$~\cite{Sedrakian:2009kb}. Let us first concentrate on the
BCS limit. One remarkable feature of the FF state is that, once it is
allowed for, the anomalies seen in Figs. \ref{fig2} and \ref{fig3} are
removed, i.e., the functional dependence of the physical quantities on
the temperature are the same as in the ordinary BCS theory. The only
effect of asymmetry is the reduction of the magnitude the
gap~\cite{Sedrakian:2009kb,He:2006vr}. Thus, the $T$-dependence of
quantities for $\alpha \neq 0$ are self-similar to the symmetrical BCS
phase. The low-temperature topology of the Fermi spheres in the FF
state is the same as in the asymmetrical BCS case, but only for
selected directions, such as along in the direction orthogonal to the
axis of symmetry breaking. This direction is chosen by the condensate
spontaneously. At other angles the appearance of the empty strip
(breach) mentioned above is suppressed. The FF phase occupies the
low-temperature and large asymmetry portion of the $T$-$\alpha$ phase
diagram and there exists a tri-critical Lifshitz point where the
normal phase, the BCS phase, and the FF phase all meet (for more
details see ref.~\cite{Sedrakian:2009kb}). Now let us comment on the
transition from the BCS-FF state to the BEC state. Generally, as one
moves towards the BEC limit, the portion of the $T$-$\alpha$ phase
diagram occupied by the FF phase becomes smaller.  A key question is
whether the FF state exists only in the weakly coupled regime of the
phase diagram or does it persist into the strongly coupled regime.
The answer is best provided by the form of the condensate
wave-function.
\begin{figure}
\begin{center}
{\includegraphics[width=9.5cm,height=7cm]{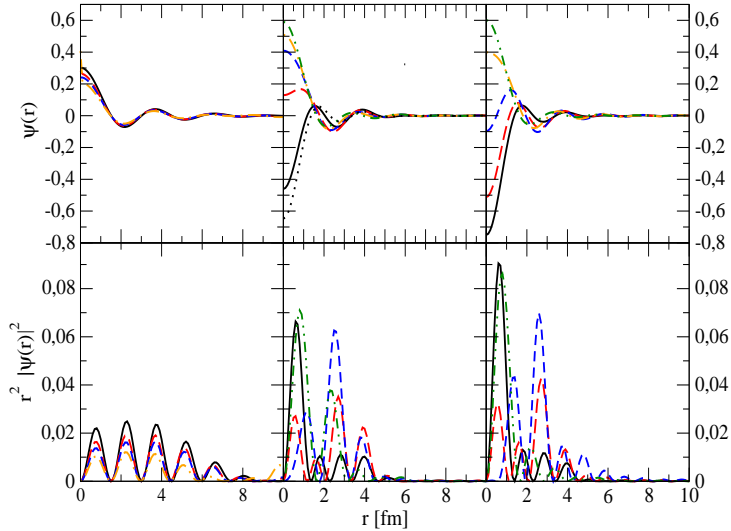}}
\caption{\label{fig6} BCS-LOFF-BEC crossover in an imbalanced system.
  The dependence of the condensate wave-function $\psi(r)$ and the
  quantity $r^2 \vert \psi(r)\vert^2$ on the spatial coordinate; the
  coupling increases from left to right. For an explanation see the
  text. }
\end{center}
\end{figure}
Fig.~\ref{fig6} displays the evolution of the condensate wave-function
from the weak (left column) to the strong (right column) coupling
regime. In the weak coupling limit the wave-function and the quantity
$r^2 \vert \psi(r)\vert^2$ show oscillations with a period given by $l
\simeq 2\pi k_F^{-1}$, where $k_F$ is the Fermi wave-number. These
sustained oscillations in space evidence the long-range correlations
characteristic to BCS state; in this regime Cooper-pair size is much
larger than the inter-particle spacing.  In the strong coupling limit
and for small asymmetries (solid line) a pronounced peak in these
quantities corresponds to a well-localized bound state. However, for
large asymmetries the picture is more complicated; there are several
peaks which are not necessarily localized at the origin. These
features indicate that a perfect BEC condensate has not been formed
yet. Since the condensate is in the FF state sector of the
$\alpha$-$T$ phase diagram (see ref.~\cite{Sedrakian:2009kb}) one may
conjecture that the FF state has a strong-coupling counterpart, where
tightly bound states carry a super-current.

\section*{Acknowledgments}
The results discussed above are based mainly  on the 
papers~\cite{LNSSS,Sedrakian:2006mt,Sedrakian:2009kb} and 
I would like to thank my colleagues for their contribution and
insight. This work was partially supported by the Deutsche
Forschungsgemeinschaft (Grant SE 1836/1-2).

\end{document}